# Comparison of Global and Local Adaptive Coordinates for Density Functional Calculations

*D. R. Hamann*

Bell Laboratories, Lucent Technologies

Murray Hill, New Jersey 07974

*ABSTRACT*

A globally-adaptive curvilinear coordinate formalism is shown to be easily convertible to a class of curvilinear transformations locally optimized around atom sites by a few parameters. Parameter transferability is established for a demanding test case, and the results of the two methods are shown to be comparable. Computational efficiencies realized in the local method are discussed.

## I. Introduction

The convergence of plane-wave pseudopotential electronic structure calculations for materials with first-row or 3-d transition metal atoms can be greatly enhanced by use of curvilinear coordinate transformations. As introduced by Gygi, the coordinate transformation was expressed in a completely general form with many parameters, and these parameters were varied along with the wave function coefficients and atomic positions to minimize the density-functional total energy.[1] This method has been shown to be highly effective for solid state calculations,[2,3] and particularly compatible with generalized-gradient approximation density functionals.[4,5,6] In Gygi's original formulation, the coordinate transformation "found" the atoms which required spatial frequencies much higher than the average plane wave cutoff through adaptation driven by energy minimization. The basis functions, plane waves as modified by the coordinate transformation, depended implicitly but not explicitly on the atomic positions, so no Pulay forces[7] on the atoms had to be considered.

The introduction of fictitious dynamics for the adaptive coordinate transformation parameters[1] proved to be compatible with quantum molecular dynamics (QMD) for the wave functions and atoms[8] and effective in calculations in which energy minimization was the goal.[2] The additional set of dynamical variables does introduce a third time scale. While ideally the fictitious masses governing these dynamics should be chosen so that the wave function variables relax rapidly compared to the adaptive coordinate variables, and those relax rapidly compared to the atomic coordinates, spanning such a range of time scales is computationally impractical. Since it is not practical to separate the coordinate and atom time scales sufficiently to decouple these systems for real-time mo-lecular dynamical studies, Gygi turned to an alternative coordinate transformation based on atom-centered local deformation functions.[9,10] A variation on this approach has been proposed by another group.[11] These studies focused on simple molecules in large supercells. Simple deformation functions were chosen, with parameters adjusted to optimize adaptive isolated atom calculations.[9]

The purpose of the present study is to compare the results of globally adaptive and atom-centered deformation-function transformations in the context of solid-state calculations, and to examine the transferability of the deformation functions. This latter point is an issue because deformation functions for neighboring atoms overlap, and the overall coordinate transformation is determined by bond lengths, density, and coordination. If the parameters of optimum deformation functions (for a given average plane-wave cutoff) vary significantly with these conditions, the deformation-function method using fixed functions will necessarily introduce systematic errors compared to the globally adaptive approach.

## II. Theoretical formulation

The local-deformation-function formalism has been described primarily for application to real-space grid calculational methods.[10,11] In addressing the issues under consideration here, it is important to utilize a formalism that has variational properties, so a plane-wave basis in curvilinear coordinate ($\xi$) space is retained as in the globally adaptive case.[1] This offers the additional advantage of keeping the calculations to be compared as similar as possible, and of minimizing additional program development. The transformations relating the two methods will be described below. In the deformation-function method, the basis set (of plane waves in the curvilinear system) is explicitly dependent on the atomic coordinates, so Pulay forces[7] must be calculated. The derivatives of the various terms in the energy with respect to the coordinate transformation were already necessary in adaptive calculations since they provided the forces which allowed the transformation to be dynamically optimized. These results from the globally adaptive case are easily converted into the required Pulay forces.[7]

In the original globally adaptive coordinate method,[1] the transformation from curvilinear to Euclidean coordinates **x** is given by

$$x_j(\xi) = \xi_j + \sum_{\mathbf{G}} x_{j,\mathbf{G}}\, e^{i\mathbf{G}\cdot\xi}, \tag{1}$$

where **G** is a set of reciprocal lattice vectors cutoff at a maximum (equivalent electron) kinetic energy $E_{Coord}$, and $x_{j,\mathbf{G}}$ is the set of adaptive parameters. In the deformation-function method, it is necessary to specify the inverse transformation, which is given by

$$\xi_j(\mathbf{x}) = x_j + \sum_{\alpha}\sum_{\mathbf{R}} (x_j - u_{j.\alpha} - R_j) f_\alpha(|\mathbf{x} - \mathbf{u}_\alpha - \mathbf{R}|) \tag{2}$$

where $\mathbf{u}_\alpha$ and $f_\alpha$ are the position and deformation function of atom α, and **R** are lattice vectors.[9] All the calculations in either form of the method are carried out on a uniform grid of points in **ξ**-space. While it is simple to calculate the corresponding (nonuniform) **x** grid using Eq. (1), this task is much more complex using Eq. (2), since a coupled set of 3 non-linear equations must be solved for each mesh point. This can be done iteratively using Newton's method. For use below, we define atom-by-atom contributions

$$\left[\frac{\partial \xi_k(\mathbf{x})}{\partial x_j}\right]_\alpha = \sum_{\mathbf{R}} \left[ \delta_{ij} f_\alpha(|\mathbf{x} - \mathbf{u}_\alpha - \mathbf{R}|) + \frac{(x_j - u_{j,\alpha} - R_j)}{|\mathbf{x} - \mathbf{u}_\alpha - \mathbf{R}|} f_\alpha{}'(|\mathbf{x} - \mathbf{u}_\alpha - \mathbf{R}|) \right], \tag{3}$$

and the summed derivative matrix

$$\frac{\partial \xi_i}{\partial x_j} = \sum_{\alpha} \left[ \frac{\partial \xi_k}{\partial x_j} \right]_{\alpha} . \tag{4}$$

Successive approximations $\mathbf{x}^n$ are generated by iterating with the inverse of the summed derivative matrix,

$$x_j^{n+1}(\xi) = x_j^n + \sum_{i=1}^{3} \left[ \frac{\partial x_j}{\partial \xi_i} \right]_{\xi=\xi(\mathbf{x}^n)} [\xi_i - \xi_i(\mathbf{x}^n)] . \tag{5}$$

Since $f_\alpha, f_\alpha'$, and their sum over the atoms must be evaluated repeatedly during these iterations, it is advantageous to keep $f$ simple and to make sure it goes to zero sufficiently rapidly at large distances. Since the coordinate transformation changes very little in one molecular dynamics step, this calculation converges very rapidly. Ramping the amplitudes of the deformation functions from zero to their desired values over the first 10 or 20 time steps yields stable starts.

Once the complete **x** grid is determined, the $\{\mathbf{x_G}\}$ parameters can be calculated by Fourier transform,

$$x_{i,\mathbf{G}} = \Omega^{-1} \int_{\text{Cell}} d^3\xi [x_i(\xi) - \xi_i] e^{-i\mathbf{G}\cdot\xi} , \tag{6}$$

where $\Omega$ is the unit cell volume, and the calculation can proceed as in the adaptive case. The $\boldsymbol{\xi}$ grid spacing is determined so that the charge density can be Fourier transformed without aliasing error. This demands that the average energy cutoff of the charge density **G** set be 4 times the average wave function basis cutoff energy $E_w$. ("Average" here refers to the energy cutoffs in $\boldsymbol{\xi}$ space. Effective cutoffs in **x** space are boosted to much higher values by the coordinate transformation.) While it would appear to be appropriate to keep the full set of $\{\mathbf{x_G}\}$ parameters so that the $\mathbf{x}$ grid determined by Eq. (5) could be reproduced exactly, the inversion of Eq. (2) can apparently produce rather high spatial frequencies even when the $f$ functions are very smooth, especially if large local enhancements of the cutoff are sought. It is advantageous to set $E_{Coord} \approx 3E_w$ to produce a modest amount of Fourier filtering. This greatly reduces aliasing errors in the evaluation of quantities such as the covariant Laplacian, which involve higher derivatives of the transformation. The filtered transformation is perfectly well defined, and no errors are induced in the subsequent evaluation of Pulay forces. Such filtering typically moves $\mathbf{x}$ grid points no more than a few parts in $10^{-3}$ of the average grid spacing. The empirical rule for $E_{Coord}$ for deformation-function based transformations is to be contrasted with the much lower cutoff $E_{Coord} \approx 0.3E_w$ used in globally adaptive calculations.[2,3] This limit was set not by aliasing problems, but by the fact that the equations of motion used to optimize $\mathbf{x_G}$ become extremely "stiff" and unstable for large $|\mathbf{G}|$.

The forces on $\mathbf{x_G}$ calculated for use in the adaptive method can be used to find the Pulay forces[7] on the atoms by application of the chain rule,

$$\begin{aligned} \frac{\partial E}{\partial u_{j,\alpha}} &= \sum_{\mathbf{G}} \sum_{i=1}^{3} \frac{\partial E}{\partial x_{i,\mathbf{G}}} \frac{\partial x_{i,\mathbf{G}}}{\partial u_{j,\alpha}} \\ &= \Omega^{-1} \int_{\text{Cell}} d^3\xi \sum_{i=1}^{3} \frac{\partial}{\partial u_{j,\alpha}} [x_i(\xi) - \xi_i] \sum_{\mathbf{G}} \frac{\partial E}{\partial x_{i,\mathbf{G}}} e^{-i\mathbf{G}\cdot\xi} , \end{aligned} \tag{7}$$

where we have substituted Eq. (6) and rearranged terms to indicate that the $\mathbf{G}$ sum should be carried out first. This inverse transform,

$$h_i(\xi) = \sum_{\mathbf{G}} \frac{\partial E}{\partial x_{i,\mathbf{G}}} e^{-i\mathbf{G}\cdot\xi}, \tag{8}$$

is a quantity already available in the original adaptive coordinate computations. The remainder of the integrand can be expressed in terms of quantities already calculated in the course of inverting the coordinate transformation,

$$\frac{\partial E}{\partial u_{j,\alpha}} = -\Omega^{-1} \int_{\text{cell}} d^3\xi \sum_{i,k=1}^{3} h_i \frac{\partial x_i}{\partial \xi_k} \left[ \frac{\partial \xi_k}{\partial x_j} \right]_{\alpha} , \tag{9}$$

where we have made use of the relation between the $x_j$ and $u_{j,\alpha}$ derivatives of $\xi(\mathbf{x})$.

## III. Computational tests and discussion

### *A. Deformation function and transferability*

A challenging problem previously studied with the globally adaptive method is the high-pressure phase transition of the stable α-quartz phase of $SiO_2$ to the metastable stishovite phase.[3] $SiO_2$ is computationally demanding because of the highly localized O valence orbitals and the many structural degrees of freedom which need to be relaxed. This particular phase transition also demanded the use of a generalized gradient density functional[5] to obtain results close to experiment, since the local density approximation gave essentially zero energy difference between the two phases. (The experimental difference is 0.54 eV/$SiO_2$.) There are major structural changes in the transformation of quartz to stishovite, with the Si coordination increasing from 4- to 6-fold, and that of O from 2- to 3-fold. In addition, the volume decreases by 40%. These features will strenuously test the transferability of deformation functions, so this system was chosen for the present study.

The deformation function $f(r)$ for the O's was chosen to be a product of an "active" deformation function and a longer ranged function to cut off tails and reduce the computational effort of evaluating the atom sums, following the general approach of Reference 10. A rational function with an explicitly finite range was used,

$$f(r) = a\frac{[1-(br)^4]^3\theta(1-br)}{1+0.5031cr^2+0.03906(cr^2)^2+0.00192(cr^2)^3}, \tag{10}$$

where $\theta$ is the unit step function. The reciprocal of the denominator is an excellent approximation to the sech function originally suggested by Gygi[9] for $0 < cr^2 < 2$, and the cutoff numerator function is very flat in the primary deformation range for typical parameters $c \approx 1$ and $b < 0.1$. The parameter $a$ determines the maximum cutoff enhancement at the atom center (for an isolated atom), while $c$ determines the local range of the deformation.[9,10] For $b \sim 0.1$, it was found to have almost no effect, and no significant interaction with the other parameters. Globally adaptive coordinate studies indicated that no cutoff enhancement was generated at the Si site, so no $f$ was introduced for Si.

A series of calculations was carried out for quartz and stishovite at lattice constants near their experimental values with fixed atomic coordinates as previously optimized in globally adaptive calculations.[3] The cutoffs $E_w$ and $E_{Coord}$ were set to 35 and 90Ry, respectively. The PBE generalized gradient functional[6] was used because it has several desirable properties compared to the PW91 functional[5] used in Reference 3. The two functionals are numerically similar over the physically important range of their parameters, and give nearly identical quartz-stishovite energy differences. Other details of these calculations are as in Reference 3.

A coarse grid of $a$ and $c$ values was searched until the approximate location of an energy minimum was located. Systematic energy calculations at 25 points of a finer $a$-$c$ grid surrounding the minimum were then fitted with cubic polynomials for both struc-tures, and more precise minima determined from the fits. Contour plots of these fits are shown in Fig. 1. Not only are the shapes and positions of the contours very similar, but the optimum parameter values indicated by the triangles are nearly coincident for the two polymorphs despite their differences in density and coordination. This is strong evidence for the transferability of the atomic deformation functions. The optimized values averaged for the two structures are $a = 0.918$ and $c = 1.127$.

An isolated atom supercell calculation provided an independent transferability test of the O deformation function, since in this case effects of overlapping deformation functions from neighbors are completely absent. Using the optimized $a$ and $c$ values, a $20a_B$ fcc supercell, two inequivalent Bloch vectors to minimize supercell interactions, and other parameters as in the above calculations, an absolute convergence error of 0.023eV (out of 431 eV) was found relative to an exact radial-mesh calculation.

*B. Comparison of results*

The globally adaptive coordinate method has been very thoroughly tested in the context of $SiO_2$.[2,3] To test the deformation-function method in a comparable fashion, a complete set of energy-volume calculations was performed for both quartz and stishovite using the average optimized $a$ and $c$ values and other parameters as above, with previously optimized internal coordinates and lattice parameter ratios.[3] Since the switch to the PBE density functional[6] and to $E_w = 35$ Ry hinders absolute comparisons of these results with those of Reference 3, a new series of globally adaptive calculations was undertaken as well. Initially, $E_{Coord} = 10$ was used for these calculations (which provided considerably more variational flexibility than $E_{Coord} = 5$ in Reference 3). It was very surprising to find that these calculations, with 70 and 15 symmetry-independent adaptive coordinate parameters for quartz and stishovite, respectively, were 0.0127 Hartree *less* well converged than the 2-parameter deformation-function calculations.[12] After some experimentation, it was found that $E_{Coord} = 20$ (yielding 195 and 38 parameters) was necessary to obtain comparably converged results. The maximum effective kinetic energy cutoffs of the adaptive basis, which are found from the determinant of the Riemannian metric tensor of the curvilinear coordinate transformation,[1] are 117 (114) Ry for the globally adaptive calculation and 123 (118) Ry for the deformation-function calculation for quartz (stishovite). A more detailed comparison of the cutoff enhancement produced by the two approaches is shown for stishovite in the contour plots of Fig. 2. The shape of the cutoff enhancement, reflecting the detailed shape of the coordinate transformations themselves, are remarkably similar for the two approaches. Given the large number of free parameters of the globally adaptive transformation, it is clear that the deformation function shape specified in Eq. (10) is quite optimal.

The globally adaptive results, the deformation function results, and fits using the Murnaghan equation of state[13] are compared in Fig. 3. For one volume near the energy minimum for each structure, the energy calculated with a conventional plane-wave code at a cutoff of 130 Ry is plotted for comparison. Absolute total energies are plotted in all cases – the different calculations are not "aligned" in any way. Discrepancies in both the quartz-stishovite energy difference and the overall E(V) relations are seen to be minimal.

Quantitative comparisons of the fits are made in Table 1. Results are included from recent all-electron linear augmented plane wave (LAPW) calculations using the PBE functional.[14] The PW91 results from Reference 3 are also included to demonstrate that their differences from the new PBE results are minimal. The largest discrepancy in the table is $B_0'$ for stishovite. The deformation function result in this case is closer to the LAPW results, and the reason for the discrepancy from both globally-adaptive coordinate results is not clear.

The relative performance of the two methods as a function of the cutoff $E_w$ is also of interest, especially compared to the conventional plane wave basis set. The simple "ideal β-cristobalite" structure[15] was used for these tests as in a previous such comparison.[2] The convergence errors for conventional plane wave, globally adaptive, and deformation function calculations are plotted in Fig. 4, where a conventional calculation with $E_w = 250\,\text{Ry}$ was used to provide the reference energy. For expediency, $E_{Coord} = 20$ was maintained for all the globally adaptive calculations, and $E_{Coord} = 90$ for the deformation-function calculations, with the charge-potential cutoff maintained $\geq 140\,\text{Ry}$ to avoid aliasing errors. (Cutoffs scaled in proportion to $E_w$ would be more computationally efficient at low $E_w$ cutoffs, but the very low cutoffs included in Fig. 4 are too poorly converged to be of practical use anyway.) The deformation function *a* and *c* parameters were kept at the optimum values found for quartz and stishovite for the 35Ry calculation, and re-optimized for 12 and 20 Ry $E_w$. For these lower-cutoff calculations, the additional variational freedom inherent in the globally adaptive method apparently adds sufficiently to the limited freedom of the plane waves to give this method a convergence advantage of approximately a factor of two. By $E_w = 35$ this advantage has disappeared, with the larger variational freedom of the plane wave basis obviating the need for more complex shapes in the coordinate transformation. This trend provides some rationalization for the similarities between the two forms of coordinate transformation shown for $E_w = 35$ in Fig. 2. The small ($5\times10^{-5}$ au in this case) shear modulus introduced to stabilize global coordinate adaptation[1,2] tends to smooth features of the coordinate transformation which are sufficiently weakly driven by the density functional total energy.

*C. Computational efficiency*

The remaining issue to consider is the relative computational efficiency of the methods. The time step used in the globally adaptive coordinate case is arrived at as a compromise based on the consideration of several things. Adequate separation of time scales for the 3 classes of dynamical variables, adequate $E_{Coord}$ to deal with the required spatial scale of the curvilinear transformation, and the need for at least marginal stability for the integration of the equations of motion dictate a combination of time step and of fictitious masses for the wave functions and coordinates. The recently proposed acceleration scheme based on wave function fictitious masses which depend on the wave vector of the wave function component[16] cannot be used to advantage in this context. After eliminating the $\mathbf{x}_\mathbf{G}$ equations of motion, however, we find that the variable fictitious mass preconditioning scheme[16] allows us to increase the time step by a factor of 2.5 for the wave function cutoffs typically employed in our $SiO_2$ calculations. The mass is simply scaled as suggested in Reference 16 with the average wave vector, that is the wave vector in $\xi$ space. Despite the fact that the effective kinetic energy of each $\xi$ plane wave is a strongly varying function of coordinate, this procedure is effective in stabilizing the equations of motion at the larger time step. For fixed atomic positions, the improvement in performance is even more dramatic since the time scale for the $\mathbf{x}_\mathbf{G}$ dynamics is re-moved from the problem altogether. The globally adaptive calculations at the energy minimum point for the quartz results in Fig. 3 took 10 times as long as those for the deformation function because of the time to converge the $\mathbf{x}_G$. Some of this additional advantage is retained even with atomic relaxation because a substantial number of iterations with active $\mathbf{x}_G$ dynamics is necessary before the forces are well enough converged to start the atom dynamics. As an additional reference point, we note that the 130 Ry ordinary plane wave quartz calculation took 3 times as long as the deformation function calculation. Since even with electronic mass preconditioning of these equations of motion the time step had to be reduced by a factor of 0.6 compared to the deformation function calculation, the overall advantage when calculating atomic dynamics is a factor of 5.

**IV. Conclusions**

In summary, we have found that despite the aesthetic appeal of employing totally unbiased curvilinear coordinate transformations based on a globally adaptive formalism, transformations based on appropriately chosen atom-centered deformation functions appear to be equally effective. In addition, they offer computational advantages in permitting significantly larger time steps. The author's primary motivation for exploring this issue was stimulated by difficulties that overlapping time scales for coordinate and atom motion engendered in attempting to utilize a form at atom dynamics modified to converge to a saddle point configuration.[17] The combination of "downhill" dynamics for the coordinate transformation with selectively "uphill" dynamics for the atoms proved unstable. The desirability of eliminating the intermediate time scale for problems in which real-time molecular dynamics results are required has already been noted.[9,10]

Table I. Comparison of Murnaghan fit parameters for globally adaptive and local deformation function PBE calculations, LAPW PBE calculations, and globally adaptive PW91 calculations. $V_0$ is the volume per $SiO_2$.

| Polymorph | | Gl.-Ad. PBE | Def. Fun. PBE | LAPW PBE | Gl.-Ad. PW91 |
|---|---|---|---|---|---|
| Quartz | $V_0$ (Å$^3$) | 39.6 | 39.9 | 39.4[a] | 39.3[b] |
| | $B_0$ (Gpa) | 45 | 42.4 | 44[a] | 48[b] |
| | $B_0'$ | 3.1 | 3.3 | 3.2[a] | 3.0[b] |
| Stishovite | $V_0$ (Å$^3$) | 24.5 | 24.4 | 24.2[a] | 24.6[b] |
| | $B_0$ (Gpa) | 237 | 246 | 257[a] | 260[b] |
| | $B_0'$ | 3.1 | 5.0 | 4.9[a] | 3.0[b] |

[a]Reference 14

[b]Reference 3

**Figure Captions**

Fig. 1. Contour plots of energy (Hartree) as a function of deformation function parameters *a* and *c* in Eq. (10) for two silica polymorphs. The energy zeros are set at the minima, and the minima are indicated by solid triangles for each polymorph. The minimum for the other polymorph is reproduced as an open triangle in each plot.

Fig. 2. Contour plots of the effective plane wave cutoff energy (Ry) for a plane in stishovite. The plane intersects O atom centers on the top and bottom edges of each plot at the positions of the cutoff maxima. The presence of Si atoms at the lower and upper left corners and on the center of the right edge of the plots has no apparent effect on the contours.

Fig. 3. Energy vs. volume curves for quartz and stishovite calculated by several methods.

Fig. 4. Convergence error as a function of average plane wave kinetic energy cutoff.

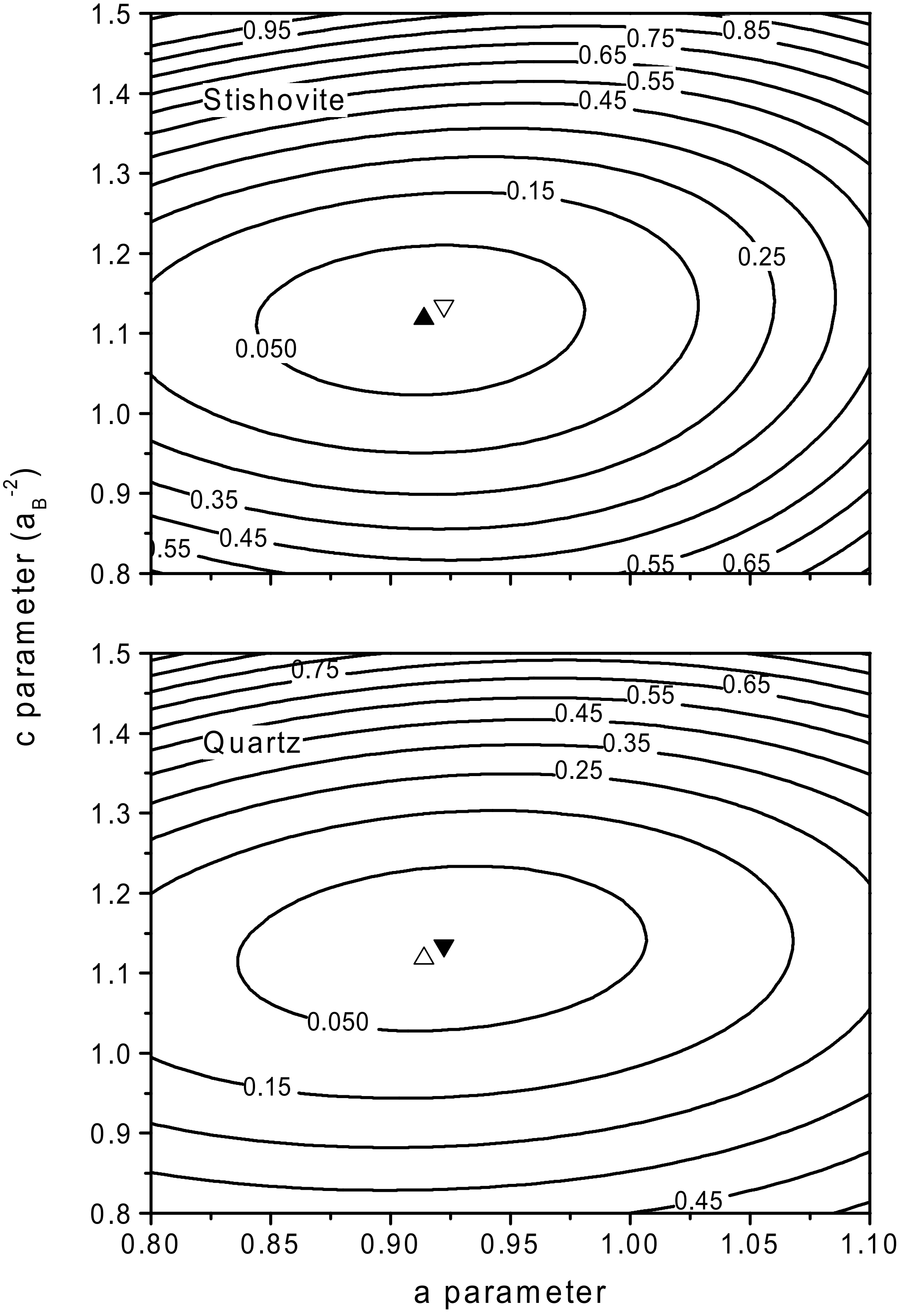

Fig. 1

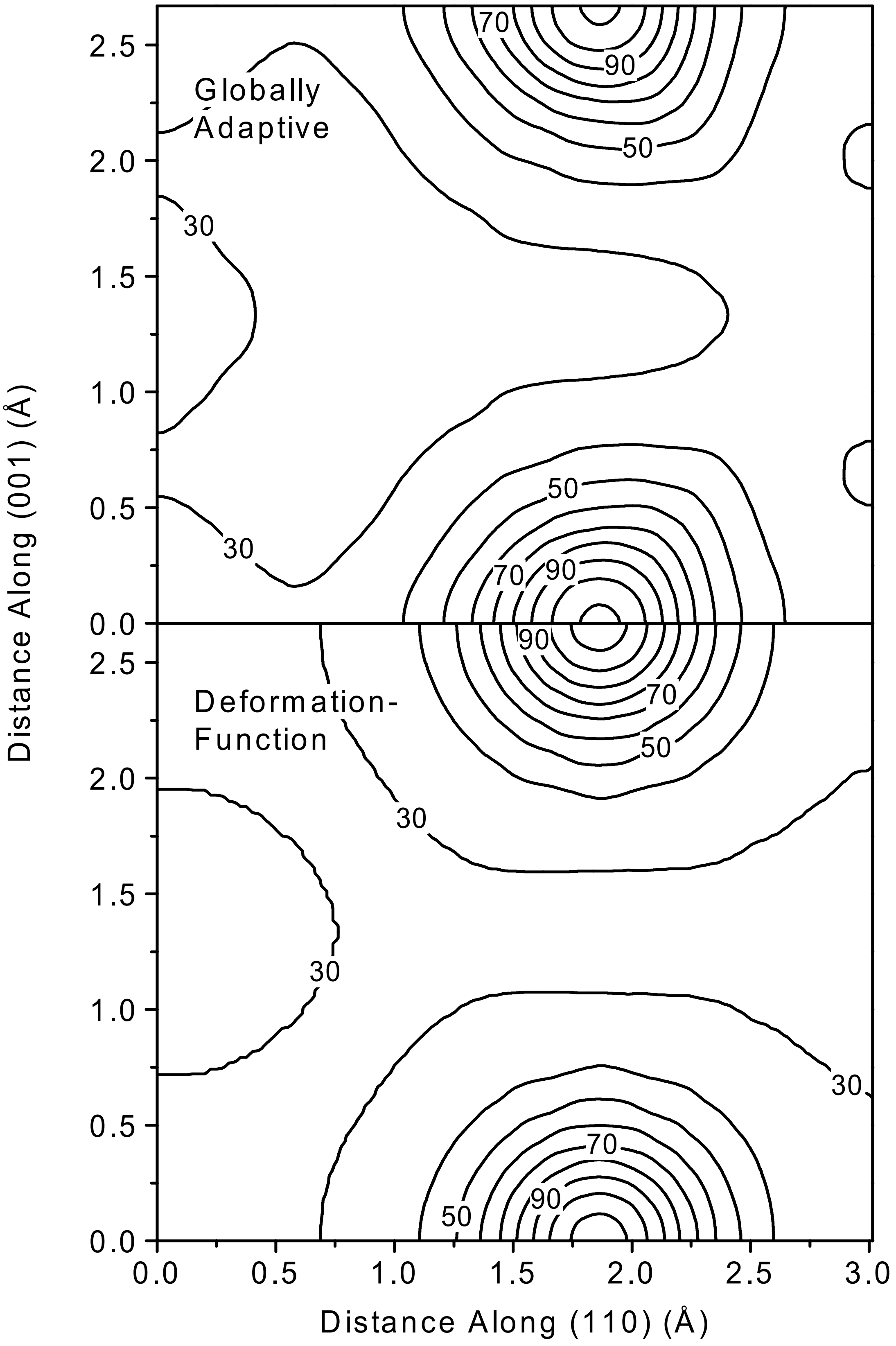

Fig. 2

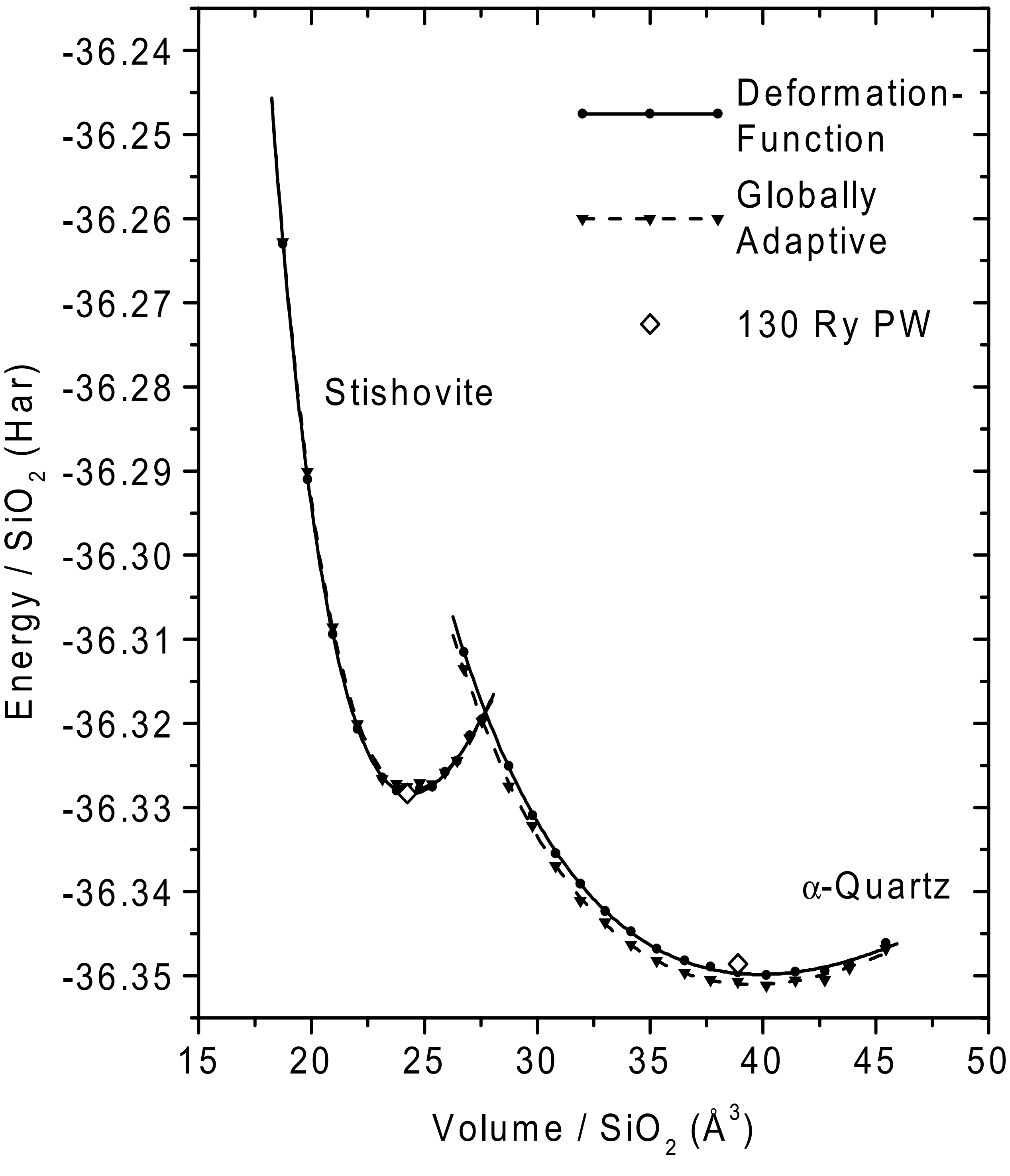

Fig. 3.

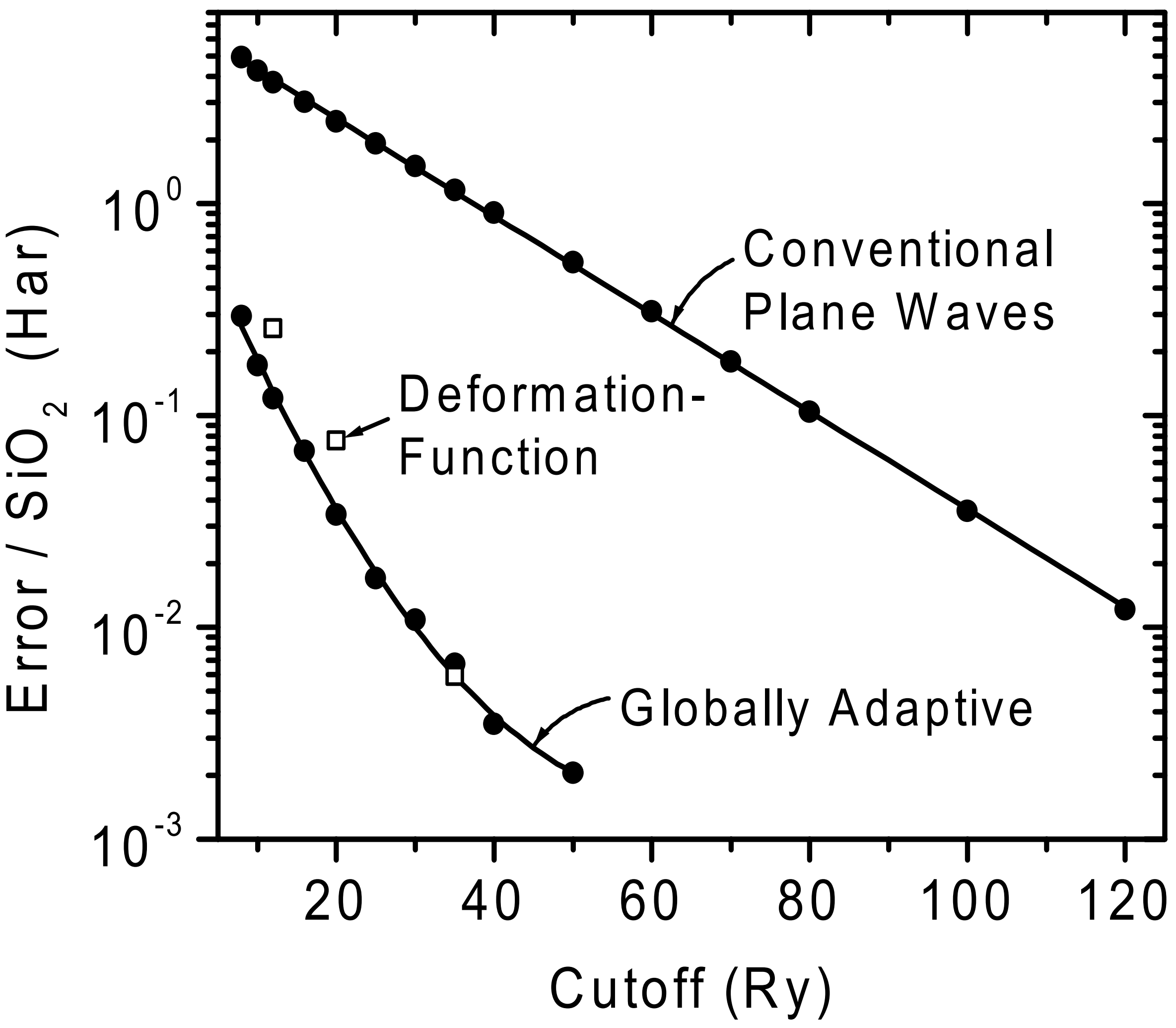

Fig. 4.